\documentclass[fleqn,twoside]{article}
\usepackage{espcrc2}

\usepackage{graphicx}
\usepackage[figuresright]{rotating}

\newcommand{\AmS}{{\protect\the\textfont2
  A\kern-.1667em\lower.5ex\hbox{M}\kern-.125emS}}
\newcommand {\pom} {I\!\!P}
\newcommand {\pomsub} {{\scriptscriptstyle \pom}}

\newcommand {\xpom} {x_{\pomsub}}
\newcommand {\apom} {\alpha_{\pomsub}}

\newcommand {\aprime} {\alpha^\prime_\pomsub}

\newcommand{\ftwodthree}{F_2^{D(3)}}

\title{Inclusive and exclusive diffraction: overview of the
  experimental presentations at Diffraction2004}

\author{Aharon Levy\address[TAU]{ Raymond and Beverly Sackler Faculty
    of Exact Sciences, School of Physics and Astronomy, Tel Aviv
    University, Tel Aviv, Israel} \thanks{Supported in part by the
    Israel Science Foundation (ISF).}}
       
\begin{document}

\begin{abstract}
  Some issues in inclusive and exclusive 
  diffractive processes are discussed.
  \vspace{1pc}
\end{abstract}

\maketitle

\section{Introduction}

There were 20 experimental presentations at this conference, all by
expert people in the field. All these talks were plenary, so there is
no point of me summarizing again the contents of each talk. Instead, I
would like to present a personal view and touch upon few selected
issues which were presented here.

\section{ What is diffraction? }

In spite of the fact that diffractive processes have a long history,
it is still not easy to give a precise and concise definition of what
one calls a diffractive reaction. It is clearly a process where, in an
exchange picture, no color is exchanged. This however includes all the
colorless particles. A further requirement is that the exchanged
messenger has the quantum number of the vacuum, which, in the Regge
picture is named the Pomeron. Elastic scattering is usually presented
as an example of a diffractive process. However, at low energies it
can proceed also through an exchange with quantum numbers different
from the vacuum, called a Reggeon. At high energies, where the
kinematics allows for a large rapidity range, it is more useful to
talk about rapidity gaps. Due to the fact that no color is exchanged,
there is a suppression of gluon radiation and therefore a rapidity gap
is produced between the two vertices of the interaction. Of course,
both the Pomeron and the Reggeon are colorless, however the rapidity
gap produced by a Reggeon is exponentially suppressed, while that of
the Pomeron remains constant. This prompted Bjorken~\cite{bj} to
define diffraction as processes in which the large rapidity gap is not
exponentially suppressed.

How practical is this definition? This we shall see in the following sections.

\section{Diffraction in inclusive processes}

\subsection{Kinematical variables}

\begin{figure}[htb]
\includegraphics[width=0.5\hsize]{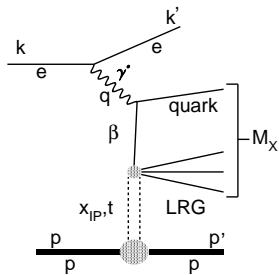}
\caption{Schematic representation of the deep inelastic diffractive 
scattering.  }
\label{fig:ddis}
\end{figure}

I will try to define here all the variables needed in the following
sections. Let us first look at a diffractive reaction $ep\to epX$ at HERA,
depicted in figure~\ref{fig:ddis}.  
A virtual photon $\gamma^*$ ($Q^2\equiv-q^2$) interacts
with the proton through a colorless exchange with vacuum quantum
numbers. A mass $M_X$ of the hadronic system recoils against the
proton. The square of the momentum transfer at the proton vertex is
$t$. The fraction of the proton momentum carried by the exchange is
denoted by $\xpom$. The quark struck by the virtual photon carries a
fraction $\beta$ of the momentum of the colorless exchange. (Note that
sometimes $z$ is used instead of $\beta$). The last two variables are
connected to Bjorken $x$ as follows: $x=\beta\xpom$.

At the Tevatron, where one studies the diffractive reaction $\bar{p} p
\to \bar{p} X$, the equivalent variable to $\xpom$ is denoted by $\xi$.

\subsection{Selecting diffractive processes}

There are three methods used at HERA to select diffractive events.
One~\cite{lps} uses the Leading Proton Spectrometer (LPS) to detect
the scattered proton and by choosing the kinematic region where the
scattered proton looses very little of its initial longitudinal
energy, it ensures that the event was diffractive. A second
method~\cite{lrg} simply request a large rapidity gap (LRG) in the
event and fits the data to contributions coming from Pomeron and
Reggeon exchange. The third method~\cite{mx} uses the distribution of
the mass of the hadronic system seen in the detector, $M_X$, to
isolate diffractive events. We will refer to these three as LPS, LRG
and $M_X$ methods. At the Fermilab Tevatron~\cite{dino} diffractive
processes are being studied by tagging events with either a rapidity
gap or a leading hadron.

The LPS method has the advantage of detecting the scattered proton and
thus excluding proton dissociative processes. However, in order to
ensure that the scattered proton resulted from a diffractive process,
one requires $\xpom <$ 0.01, where $\xpom$ is the amount of
longitudinal momentum lost by the scattered proton. This cut removes
contributions coming from Reggeon exchanges~\cite{xpcut}.

The LRG method selects events which also include some proton
dissociative processes and Reggeon contributions. The latter can be
removed by the same $\xpom <$ 0.01 cut as above. The proton
dissociative processes can be removed provided their mass is large
enough to produce signals in some forward tagging devices. The
contribution of low mass proton dissociation can be estimated.  In the
analysis of the H1 collaboration, processes with proton dissociation
into masses below 1.6 GeV amount to about 10\%~\cite{sebastian}.

The $M_X$ method which subtracts the exponentially suppressed large
rapidity gap events, in principle subtracts also the Reggeon
contribution and is left only with the proton dissociative background.
These can not be removed for masses below 2.4 GeV, which constitutes
about 30\% of the selected diffractive events~\cite{capua}.

At the Tevatron, single diffractive events, $\bar{p} p \to \bar{p} X$,
are selected by tagging the scattered $\bar{p}$.

\subsection{Diffractive structure function}

\begin{figure}[htb]
\includegraphics[width=1.0\hsize]{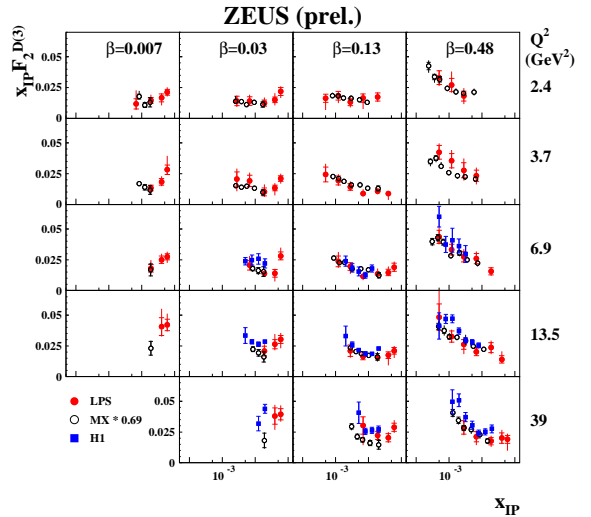}
\caption{Comparison of $\xpom\ftwodthree$ measured by H1 and ZEUS, as a 
function of $\xpom$ in overlapping bins of $\beta$ and $Q^2$. }
\label{fig:allsf}
\end{figure}

In figure~\ref{fig:allsf} the diffractive structure function
measurements with all three HERA methods are presented. The $M_X$ data
have been multiplied by a factor of 0.69 to correct for the proton
dissociation background. No correction was done to the H1 data. All
methods seem in general to agree with each other in their overlapping
kinematic region.

\subsection{$Q^2$ dependence of $\lambda$}

The $x$ behaviour of the inclusive structure function $F_2$ at a given
$Q^2$ is well described by an $x^{-\lambda}$ form. The value of
$\lambda$ is connected to the Pomeron intercept, $\lambda=\apom(0)-1$.
The value of $\lambda$ is approximately constant till $Q^2\approx$ 1
GeV$^2$, and then rises almost linearly with $\ln Q^2$.

It is of interest to see whether the $\xpom$ behaviour of $\xpom
\ftwodthree$ shows a similar pattern. To this end, a fit of the form
$\xpom \ftwodthree \propto \xpom^{-\lambda}$ was performed, for different
$Q^2$ intervals.

\begin{figure}[htb]
\includegraphics[width=1.0\hsize]{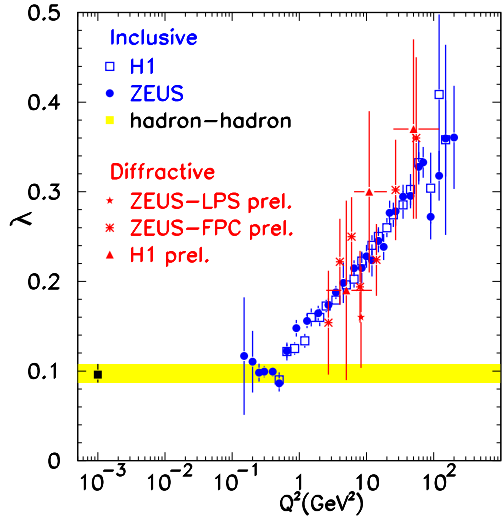}
\caption{The $Q^2$ dependence of $\lambda$ obtained from fits to 
$F_2\propto x^{-\lambda}$ and $\xpom \ftwodthree \propto \xpom^{-\lambda}$. 
The band indicates the value corresponding to the soft Pomeron.  }
\label{fig:lama}
\end{figure}

Figure~\ref{fig:lama} shows the value of $\lambda$ as function of $Q^2$
from fits to the $x$ behaviour of $F_2$ and from the $\xpom$ behaviour
of $\xpom \ftwodthree$. The precision data of $F_2$ makes it possible
to see a very significant rise with $Q^2$. The $\xpom \ftwodthree$ data
does not have the precision needed for a clear $Q^2$ dependence. There
is a trend similar to that of $F_2$, but given the large errors of the
data, the behaviour is also consistent with no $Q^2$ dependence.

\subsection{NLO QCD fits to $\xpom \ftwodthree$}

It has been proven~\cite{QCD-fact} that QCD factorization works for
diffractive processes at HERA. This allows to use the
DGLAP~\cite{dglap} evolution equations to get diffractive
parton distribution functions. Given the fact that for describing
diffractive processes one needs more variables, $t$, $\xpom$, $\beta$,
$Q^2$, one would actually like to evolve in $\beta$ and $Q^2$ for
fixed $t$ and $\xpom$. $t$ is usually hard to measure and one
integrates over it. Thus, ideally one would like to evolve for fixed
$\xpom$ values. The statistics of the presently available data is not
sufficient for carrying this out.

Ingelman and Schlein~\cite{ingelman-schlein} suggested to consider the
exchanged Pomeron as a particle having internal structure. Under this
assumption, the diffractive process is described as a multistep event:
the proton 'radiates' a Pomeron having a fraction $\xpom$ of the
proton momentum. The virtual photon interacts in a deep inelastic
process off the Pomeron, scattering of a parton in the Pomeron
carrying a fraction $\beta$ of the Pomeron momentum. There is thus a
flux factor at the proton vertex, dependent only on $\xpom$ ($t$ is
integrated out), and a Pomeron structure function $F_2^{D(2)}$. This
picture assumes Regge factorization, an assumption which has to be
checked by the data.

Using the proven QCD factorization together with the assumed Regge
factorization, one gets diffractive parton distributions.

\subsection{Regge factorization}

The assumption of Regge factorization clearly does not hold in the
inclusive case. The value of $\lambda$ is clearly $Q^2$ dependent.
However both the diffractive H1 data and the LPS data can be described
by an NLO QCD fit with one fixed value of $\apom$(0), as shown by
Sch\"atzel~\cite{sebastian} and by Capua~\cite{capua} at this meeting.
In case of the LPS data, a cut on $\xpom <$ 0.01 has to be made. For
the H1 analysis, one needs to add the Reggeon contribution. The
statistics of the LPS data were not sufficient to repeat the fit in
different $Q^2$ bins. The H1 analysis, shows some indication of a rise
of $\lambda$ with $Q^2$ (see figure~\ref{fig:lama}), though with quite
large error bars. The uncertainty comes not only from the statistics
but also from the way one needs to isolate the Pomeron contribution
from the Pomeron + Reggeon fit.

We can conclude that in inclusive diffractive processes, for $\xpom <$
0.01, the data can be consistent with Regge factorization. There is a
clear need for more precise data.

\subsection{Diffractive parton distribution functions}

\begin{figure}[htb]
\includegraphics[width=1.0\hsize]{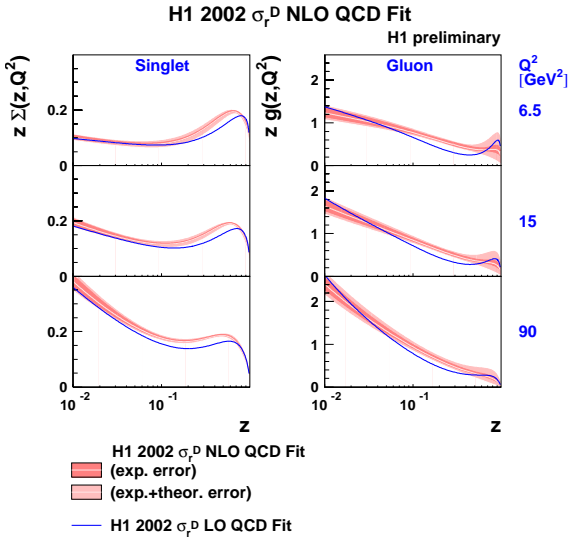}
\caption{The resulting parton density distributions in the Pomeron, using a 
NLO QCD fit (shaded line) compared to a LO fit (solid line).}
\label{fig:dpdfs}
\end{figure}

\begin{figure}[htb]
\includegraphics[width=0.6\hsize,clip=true]{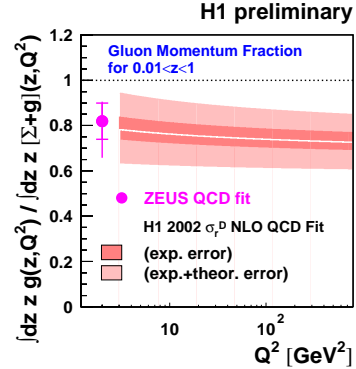}
\caption{The gluon momentum fraction from a NLO QCD fit, as a function of 
$Q^2$. }
\label{fig:gluon-fract}
\end{figure}

The H1 data, which have a wide kinematical coverage in $Q^2$ and in
$\beta$, have been used to extract diffractive parton distributions
(dpdfs), shown in figure~\ref{fig:dpdfs}.
One sees a dominance of the gluon
distribution, which is the outcome of the fact that the data show
positive scaling violation up to quite high $\beta$ values. This is
quantified in figure~\ref{fig:gluon-fract}, which shows that for the
region $0.01 < \beta < 1$, the gluons carry 80\% of the Pomeron momentum. 
The same conclusion is reached by the LPS analysis. Note that the
validity of the diffractive parton distribution functions (dpdfs) is
in the following kinematic region: $Q^2 >$ 3 GeV$^2$, $M_X >$ 2 GeV
and $\xpom <$ 0.05. Are these dpdfs portable just like in the
inclusive pdfs case? In other words, does QCD factorization work?

\subsection{QCD factorization test}

The diffractive parton distributions, obtained by H1 from the NLO QCD
fit, were used to calculate expectations of other diffractive
processes. The agreement with data is very good for diffractive $D^*$
as well as for diffractive dijet production~\cite{vinokurova}.
However, the expectations for the Tevatron results are by one order of
magnitude too high~\cite{mesropian}. This is not surprising as QCD
factorization should not hold for diffractive hadron-hadron reactions.
Furthermore, the Tevatron data is measured in the kinematic region
$0.035 < \xi <0.095$, where the Reggeon exchange dominates and thus
would not be called diffraction. As mentioned above, the validity of
the H1 fit is for $\xpom <$ 0.05 ($\xpom$ at HERA is $\xi$ at the
Tevatron).  The fact that QCD factorization seems to fail for
hadron-hadron data is also explained by introducing the notion of
survival probability of the rapidity gap~\cite{survival}. Taking it at
face value, this would mean that for the Tevatron processes, the
survival probability is about 0.1. Since the photon has a hadronic
part ('resolved photon'), this notion can be tested in diffractive
photoproduction of dijets~\cite{renner,vinokurova}. Indeed it seems
that using a survival probability of 0.34~\cite{kaidalov-survival},
one can describe the resolved photon data. There seems to be some
uncertainty about the conclusion concerning the direct part: while the
H1 measurement is below the expectations, the ZEUS result is
consistent with it.

\begin{figure}[htb]
\includegraphics[width=0.9\hsize,clip=true]{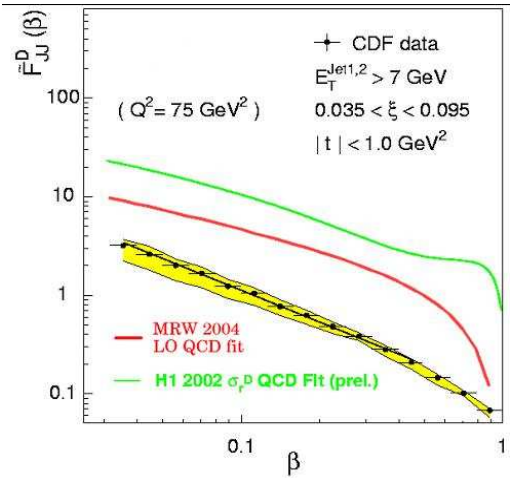}
\caption{Effective diffractive structure function for dijet production in 
$p\bar{p}$ interactions as a function of $\beta$, compared to expectations 
of different sets of diffractive parton distribution functions. }
\label{fig:watt}
\end{figure}

An interesting attempt to fit the combined data of inclusive and
inclusive diffraction data was carried out by Martin, Ryskin and
Watt~\cite{mrw}, who included absorption correction to the QCD
analysis. While the quark distributions they get are not very
different from those of H1, their gluon distribution is significantly
lower than the H1 one. This results in expectations which are by
almost factor of 3 lower than the H1 ones for the comparison with the
Tevatron data (see figure~\ref{fig:watt}).
In this case the survival probability would be closer to that of the
resolved photon case.

\subsection{Ratio of $\sigma^D$ to $\sigma_{tot}$}

\begin{figure}[htb]
\includegraphics[width=0.8\hsize]{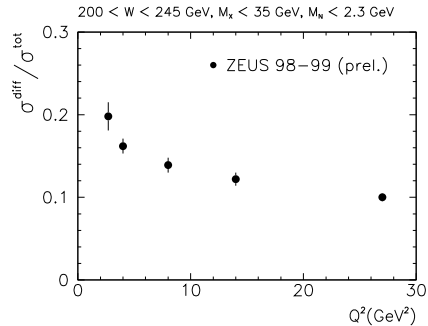}
\caption{The ratio of diffractive to total cross section as a function of 
$Q^2$, for a selected kinematic region.  }
\label{fig:sigratio}
\end{figure}

Diffractive processes were said to constitute about 10\% of the total
inclusive DIS processes. However, the ratio of
$\sigma^D/\sigma^{tot}$ is $Q^2$ dependent. It can be as high as 20\%
at $Q^2 \approx$ 3 GeV$^2$, going down to about 10\% at $Q^2 \approx$
30 GeV$^2$, as can be seen in fig~\ref{fig:sigratio}, where this ratio
is shown as a function of $Q^2$, for the kinematic region $ 200 < W <
245$ GeV, $M_X < 35$ GeV, and $M_N < 2.3$ GeV. 
One should however keep in mind that the $Q^2$ dependence might be an
outcome of the fact that in this presentation of the ratio, different
regions of $\beta$ are covered for different values of $Q^2$.  Note
that this ratio at $Q^2$ = 100 GeV$^2$ goes down to $\approx$ 5\% for
$\xpom <$ 0.03~\cite{sebastian}.

This ratio has been measured for the first time for Charged Current
diffractive processes~\cite{sebastian,rautenberg}. It is in the range
of 2-3 \% for $\xpom$ and $x <$ 0.05.

\begin{figure}[htb]
\includegraphics[width=0.8\hsize]{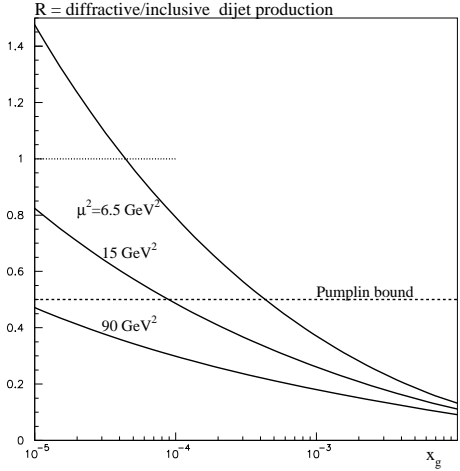}
\caption{The ratio of diffractive to inclusive dijet production cross section 
  as a function of $x$ of the gluon for different scales of the hard
  scattering, for the recent H1 diffractive parton distribution
  functions. Also shown is the unitarity limit, called Pumplin bound.
}
\label{fig:unitarity}
\end{figure}

One can calculate the ratio of diffractive to total cross section for
specific processes and check whether the parton distributions obtained
from the NLO QCD analysis fulfill the Pumplin bound~\cite{pumplin}.
This was done in~\cite{kaidalov-survival} for diffractive to inclusive
dijet production induced by gluons and is displayed in
figure~\ref{fig:unitarity}.
As seen, the bound is clearly violated for relatively low scales at
low $x$. This might indicate that unitarity effects are already
present in the gluon sector.

\subsection{Summary on inclusive diffraction}

This subsection is more a presentation of some questions than a real
summary. Is Regge factorization broken? Also for $\xpom <$ 0.01? There
is a need for more precise measurements to come to a clear conclusion.
Ideally, so as not to be dependent on the Regge factorization
assumption in diffraction, one would like to do a QCD analysis for
fixed values of $\xpom$. This will need much higher statistics than
presently available.

There is a breaking of QCD factorization when using the parton
distribution densities to compare to hadron-hadron data. This is
interpreted by the introduction of the large rapidity gap survival
probability. The value of the survival probability seems to be in the
range of 0.1-0.3.

The presently obtained gluon momentum densities seem to give results
which are violating the Pumplin bound, in certain kinematical regions.
This could be the indication of the presence of unitarity effects.

There is a large ratio of diffractive to total cross section
which decreases with $Q^2$.

\section{Exclusive diffractive processes}

\subsection{Introduction}

This section describes exclusive processes like electroproduction of
vector mesons or Deeply Virtual Compton Scattering (DVCS). The
situation in this cases is much simpler as these processes are clearly
diffractive processes at the high energies where they are measured.
There still exist the problem of isolating the 'elastic' process from
the proton dissociative one. By measuring the cross section for a
limited low $t$ range, the contribution of the latter is minimized.

\subsection{Soft to hard transition}

One of the nice features seen in these data is the transition from
soft to hard processes as one increases the scale. 
\begin{figure}[htb]
\includegraphics[width=0.95\hsize]{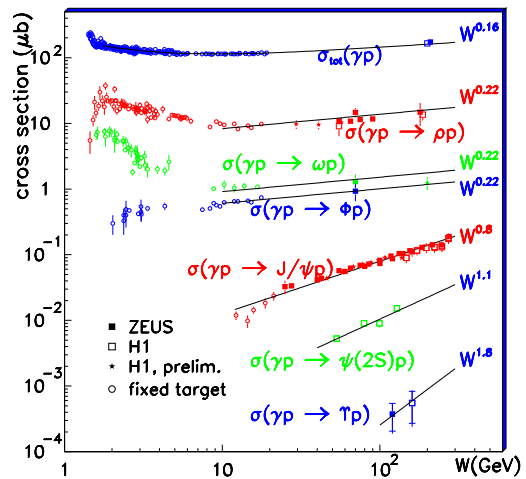}
\caption{A compilation of elastic photoproduction of vector mesons, as a
  function of $W$. The total $\gamma p$ cross section is plotted for
  comparison. }
\label{fig:xsect}
\end{figure}
This transition is
seen also for the photoproduction of vector mesons, where the mass
serves as the scale. 
Figure~\ref{fig:xsect} shows the photoproduction
cross section as a function of the $\gamma p$ center of mass energy,
$W$, for different vector mesons. 
The light vector mesons, $\rho$,
$\omega$ and $\Phi$ show an energy dependence which is characteristic
of a soft process (the total $\gamma p$ cross section is also shown
for comparison). 
For the heavier vector mesons, the energy dependence
becomes much steeper, as expected from hard processes. Note also that
the real part of the amplitude increases with the hardness of the
process and is a further reason for the sharp energy dependence.

\begin{figure}[htb]
\hspace{-0.2cm}
\includegraphics[width=0.72\hsize,angle=270]{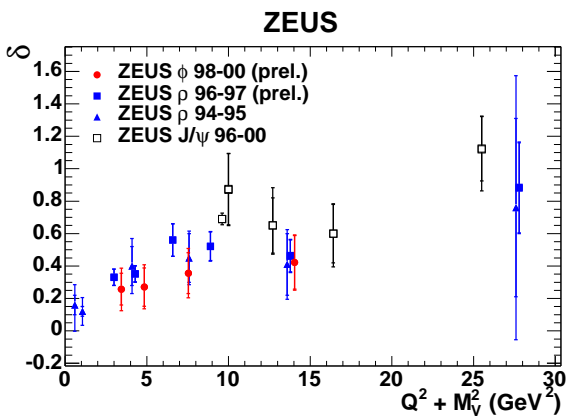}
\caption{The parameter $\delta$ from a fit of the form $W^\delta$ to the 
cross section data of exclusive production of VMs, as a function of 
$Q^2+M_V^2$. }
\label{fig:delvm}
\end{figure}

The soft to hard transition can also be seen for a given vector meson,
by changing the $Q^2$ of the process. The cross section is
parameterized as $W^\delta$ and $\delta$ is seen to increase with
$Q^2$. To compare all the vector mesons on one plot~\cite{ciesielski}
one shows $\delta$ as function of $Q^2+M_V^2$, with $M_V$ being the
mass of the vector meson. As seen in figure~\ref{fig:delvm}, one gets
an increase of $\delta$ from values of about 0.2 (soft) at the low
scale end to a value of about 1 (hard) at high scales.

\subsection{Effective Pomeron trajectory}

\begin{figure}[htb]
\includegraphics[width=1.0\hsize]{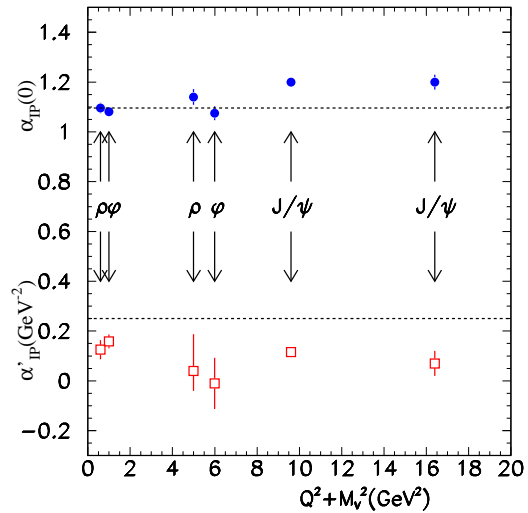}
\caption{Compilation of $\apom$(0) and $\aprime$ values, 
  extracted in exclusive VM production, as a function of $Q^2+M_V^2$.}
\label{fig:apom}
\end{figure}

Using the energy dependence of vector meson electroproduction at fixed
$t$ values, one can obtain the parameters of the effective trajectory
exchanged in the process. This way the effective trajectory of the
Pomeron was determined for the $\rho$, $\phi$ and $J/\psi$
electroproduction. A summary plot of the intercepts and slopes for all
three VMs, as function of $Q^2+M_V^2$, is presented in figure~\ref{fig:apom}.
The intercept of the low mass VMs are consistent with that of the soft
Pomeron. This is not the case for the $J/\psi$ which has a
significantly higher intercept. As for the slope, all values are lower
that that of the soft Pomeron, as expected from hard processes.

\subsection{Sizes of vector mesons}

\begin{figure}[htb]
\includegraphics[width=1.0\hsize]{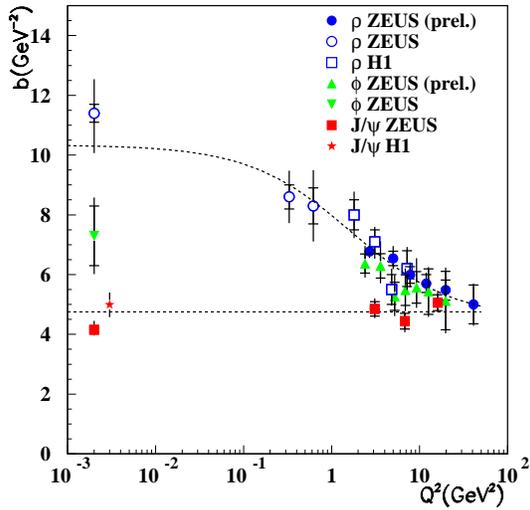}
\caption{Exponential slope of the $t$ distribution measured for exclusive 
VM production as a function of $Q^2$. The lines are to guide the eye. }
\label{fig:bvm}
\end{figure}

The $t$ distribution of the VMs can be well described by an
exponentially falling cross section, with a slope $b$. This slope is
connected to the size of the VM. At low $Q^2$ the size of the light
VMs is large, decreasing with $Q^2$ from a value of $\approx$ 10
GeV$^{-2}$ to about 5 GeV$^{-2}$. For the $J/\psi$ the size is small
already at low $Q^2$. This can be seen in figure~\ref{fig:bvm}, where
the slope $b$ is plotted as function of $Q^2$.
The dotted lines are just to guide the eye. The data seem to converge
at high $Q^2$ on a value of $b \approx$ 5 GeV$^{-2}$.

\subsection{$R=\sigma_L/\sigma_T$}

The ratio $R$ of the longitudinal photon cross section, $\sigma_L$, to
that of the transverse photon, $\sigma_T$, can be obtained by studying
the decay distribution of the VM and assuming s-channel helicity
conservation. This has been done for
$\rho$~\cite{sandacz,grebenyuk,ciesielski}, $\phi$~\cite{ciesielski}
and $J/\psi$~\cite{ciesielski}. 
\begin{figure}[htb]
\includegraphics[width=0.8\hsize]{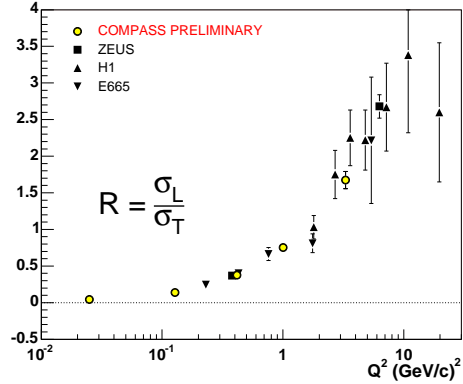}
\caption{The $Q^2$ dependence of $R$ for exclusive $\rho^0$ 
electroproduction. }
\label{fig:r-compass}
\end{figure}
Figure~\ref{fig:r-compass} shows as an example the very impressive
preliminary measurements of $R$ as a function of $Q^2$ for the $\rho$
VM by the COMPASS collaboration~\cite{sandacz}. For all VMs, $R$ is
rising with $Q^2$. 
\begin{figure}[htb]
\includegraphics[width=1.0\hsize]{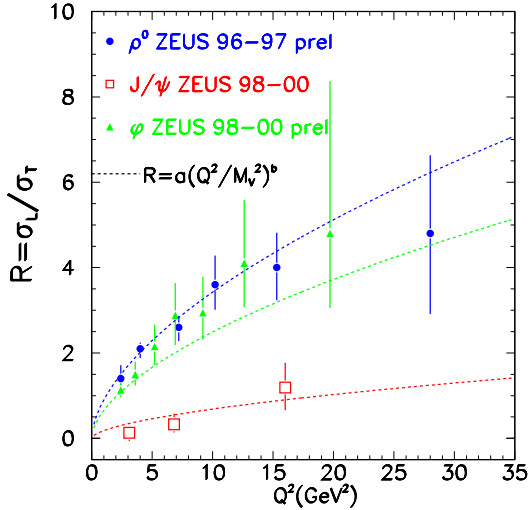}
\caption{A compilation of the values of $R$ for $\rho^0$, $\phi$ and 
$J/\psi$, as a function of $Q^2$.The lines are a fit to the
data of the form $R=a(Q^2/M_V^2)^b$. }
\label{fig:rvm}
\end{figure}
A compilation of $R$ for all three VMs is shown in
figure~\ref{fig:rvm}, as function of $Q^2$. The lines are a fit to the
data of the form $R=a(Q^2/M_V^2)^b$, which indicate that $R$ scales
with $Q^2/M_V^2$.

\subsection{Configurations of the photon}

The photon is described as fluctuating into a $q\bar{q}$ pair. When
the relative $k_T$ between the pair is small we speak of a large
spatial configuration, while if the relative $k_T$ is large, the
photon fluctuates into a small spatial configuration. Longitudinal
photons have small configurations, while transverse photons consist of
both large and small configuration. A small spatial configuration
leads to hard processes and thus to a steep energy dependence of the
cross section. A large configuration has a shallower energy
dependence, as expected in soft processes.

It came therefore as a surprise that the ratio $R$ for the $\rho$
electroproduction process is $W$ independent for $W$ values up to 120
GeV and $Q^2$ up to 19 GeV$^2$~\cite{halina}. The ratio $R$ is $W$
independent also for $J/\psi$~\cite{ciesielski}.

This means that for some reason in case of VM electroproduction the
large configurations of the transverse photon are suppressed.

Another process which seems to send the same message is DVCS. The
energy dependence of this reaction indicates a hard process. The
$W^\delta$ dependence yields~\cite{hiller} $\delta$=0.98$\pm$0.44 for
the H1 measurement and $\delta$=0.75$\pm$0.15$^{+0.08}_{-0.06}$ for
ZEUS. Such a steep energy dependence would be expected from a dominant
longitudinal photon. However, in DVCS one goes from a virtual photon
to a real one, $\gamma^*\to\gamma$. Assuming s-channel helicity
conservation, this means that, since the real photon is transverse,
also the initial virtual photon has to be transverse. Thus, the steep
energy dependence of the DVCS cross section means that the large
configuration in the transverse photon is suppressed.

\subsection{Summary on exclusive diffraction}

Exclusive diffractive processes become hard once the scale gets large.
The properties of the effective Pomeron exchange at the larger scales
are consistent with that of a hard process. The large configurations
of the transverse photon seem to be suppressed in exclusive VM,
including real photon, production.

\subsection{Outlook}

Diffractive processes are expected to be measured with improved
machines and detectors at HERA, the Tevatron and at RHIC. Whether we
want it or not, a large portion of the interactions measured at LHC
will be of diffractive nature~\cite{white}. In fact, the exclusive
diffractive production of the Higgs boson has been proposed as a
potential background free method to search for the light Higgs at LHC.
One can also dream about a future $ep$ collider which will allow to
reach kinematic region where phase transitions can be observed.
Clearly diffraction is a subject which will occupy us for quite some
time to come.

\section*{Acknowledgments}

It is a pleasure to acknowledge and thank the organizers for the
excellent organization of a very pleasant conference in a most
beautiful location.

\end{document}